\documentstyle[preprint,aps]{revtex}
\begin{document}
\draft
\title{Exact expression of the ground state energy of quantum many-particle systems
as a functional of the particle density}
\author{Yu-Liang Liu}
\address{Center for Advanced Study, Tsinghua University, Beijing 100084, \\
People's Republic of China}
\maketitle

\begin{abstract}
By introducing a phase field and solving the eigen-functional equation of
particles, we obtain the exact expressions of the ground state energy as a
functional of the particle density for interacting electron/boson systems,
and a two-dimensional electron gas under an external magnetic field,
respectively. With the eigen-functionals of the particles, we can construct
the ground state wave-function of the systems. Moreover, with the
expressions of the ground state energy, we can exactly determine the ground
state energy and the ground state particle density of the systems by taking $%
\delta E_g[\rho ]/\delta \rho (x)=0$.
\end{abstract}


\vspace{1cm}

\newpage Quantum many-particle systems are the main topics of the condensed
matter physics, in which strongly correlated electron systems are the most
interesting and hard problems, such as heavy fermion systems, high Tc
cuprate superconductors, fractional quantum Hall effects, and some
one-dimensional interacting fermion systems. In general, the quantum
many-particle (fermion) systems can be divided into two categories, one is
represented by the Landau Fermi liquid theory\cite{1,2}, and may be called
as weakly correlated systems, and another one is represented by non-Fermi
liquid theory, such as the Tomonaga-Luttinger liquid theory\cite{3,4,5} and
marginal Fermi liquid theory\cite{6,7}, and may be called as strongly
correlated fermion systems. Therefore, there exists a key parameter hidden
in the quantum many-particle systems, which represents the fermion/boson
correlation strength. In Ref.\cite{8}, we found this parameter we called the
phase field, and with it we can unifiably represent the weakly and strongly
correlated fermion systems, thus the Landau Fermi liquid theory and the
non-Fermi liquid theory can be unified under our eigen-functional
bosonization theory.

According to the Hohenberg-Kohn theorem\cite{9}, the ground state energy of
the quantum many-particle systems is uniquely determined by their ground
state particle density. This theorem does not tell us how to construct the
ground state energy by the particle density, but it clearly tells us that
the ground state energy can be exactly represented by the particle density.
However, in the Kohn-Sham scheme\cite{10,11}, we can only approximately
obtain the expression of the ground state energy as a functional of the
ground state particle density. Thus it is very important to write out an
exact expression of the ground state energy of the quantum many-particle
systems as a functional of the particle density. In fact, in usual
bosonization representation of one-dimensional interacting fermion systems%
\cite{12,13}, we learn how to construct the Hamiltonian by the fermion
density operators. With the eigen-functional bosonization theory, we can
easily treat any quantum many-particle system by using the fermion/boson
density field to represent the kinetic energy of the systems. In Ref.\cite{8}%
, we demonstrated how to unifiably represent the weakly and strongly
correlated fermion systems, and how to calculate their correlation functions.

In this paper, we give exact expressions of the ground state energy as the
particle density field for different quantum many-particle systems. These
expressions are universal, they are valid not only for weak fermion/boson
interactions, but also for strong fermion/boson interactions. We first give
the exact expression of the ground state energy as a functional of the
electron density field for interacting electron systems; secondly we give
the exact expression of the ground state energy for interacting boson
systems, and their ground state wave-functions that are very similar to the
correlated basis functions used in the study of the liquid $^{4}He$; finally
we give the exact expression of the ground state energy of two-dimensional
electron gas under an external magnetic field, and its ground state
wave-functions that have a little similarity with Laughlin's trial
wave-functions\cite{14,15}. This expression of the ground state energy has
more advantages comparing with the Laughlin's trial wave-functions, because
it directly derives from the microscopic theory.

In general, we consider the interacting electron system described by the
Hamiltonian, 
\begin{equation}
H=\psi ^{\dagger }(x)\left( \frac{\hat{p}^2}{2m}-\mu \right) \psi (x)+\frac 1%
2\int d^Dyv(x-y)\rho (y)\rho (x)  \label{1}
\end{equation}
where $\hat{p}=-i\hbar {\bf \nabla }$, $\rho (x)=\psi ^{\dagger }(x)\psi (x)$
is the electron density operator, and $D$ the dimensions of the system. In
order to calculate the ground state energy, we introduce a Lagrangian
multiplier (boson) field $\phi (x)$ which takes $\rho (x)=\psi ^{\dagger
}(x)\psi (x)$ as a constraint condition. The Hamiltonian (\ref{1}) can be
re-written as, 
\begin{eqnarray}
H[\phi ,\rho ] &=&\displaystyle{\ \psi ^{\dagger }(x)\left( \frac{\hat{p}^2}{%
2m}-\mu +\phi (x)\right) \psi }  \nonumber \\
&-&\displaystyle{\ \phi (x)\rho (x)+\frac 12\int d^Dyv(x-y)\rho (y)\rho (x)}
\label{2}
\end{eqnarray}
where $\phi (x)$ and $\rho (x)$ are independent boson fields. Analogy to the
eigen-functional bosonization theory\cite{8}, here we solve the following
eigen-functional equation, 
\begin{equation}
\left( \frac{\hat{p}^2}{2m}-\mu +\phi (x)\right) \Psi _k(x,[\phi ])=E_k[\phi
]\Psi _k(x,[\phi ])  \label{3}
\end{equation}
Using the Helmann-Feynman theorem, we have the expression of the
eigen-values, 
\begin{eqnarray}
E_k[\phi ] &=&\epsilon _k+\Sigma _k[\phi ]  \nonumber \\
\Sigma _k[\phi ] &=&\displaystyle{\ \int_0^1d\xi \int d^Dx\phi (x)|\Psi
_k(x,[\xi \phi ])|^2}  \label{4}
\end{eqnarray}
where $\epsilon _k=(\hbar k)^2/(2m)-\mu $. The eigen-functionals can be
written as, 
\begin{equation}
\Psi _k(x,[\xi \phi ])=\frac{A_k}{L^{D/2}}e^{i{\bf k}\cdot {\bf x}%
}e^{Q_k(x,\xi )}  \label{5}
\end{equation}
where $A_k$ is the normalization constant, and the phase (boson) field $%
Q_k(x,\xi )$ satisfies usual Eikonal equation, 
\begin{equation}
\left( \frac{\hat{p}^2}{2m}+\frac \hbar m{\bf k}\cdot \hat{p}\right)
Q_k(x,\xi )+\frac{[\hat{p}Q_k(x,\xi )]^2}{2m}+\xi \phi (x)=0  \label{6}
\end{equation}
It is noted that the eigen-functionals are composed of two parts, one
represents the free electron, and another one represents the contributions
from other electrons by the interaction potential $v(x-y)$, therefore the
eigen-functionals are the eigen-wave-functions of the interacting electrons
corresponding to the definite boson field $\phi \left( x\right) $. With
them, we can construct the ground state wave-function, and calculate the
correlation functions of the system by taking a functional average over the
boson field $\phi \left( x\right) $.

The electron operators $\psi ^{\dagger }(x)$ and $\psi (x)$ can be
represented as, 
\begin{eqnarray}
\psi ^{\dagger }(x) &=&\displaystyle{\sum_k\Psi _k^{*}(x,[\phi ])\hat{c}%
_k^{\dagger }}  \nonumber \\
\psi (x) &=&\displaystyle{\sum_k\Psi _k(x,[\phi ])\hat{c}_k}  \label{7}
\end{eqnarray}
where $\hat{c}_k^{\dagger }$ ($\hat{c}_k$) is the electron's creation (
annihilation) operator with momentum $\hbar k$. With equations. (\ref{3})
and (\ref{7}), we can obtain the exact expression of ground state energy as
a functional of the boson field $\phi (x)$ and the electron density $\rho
(x) $, 
\begin{eqnarray}
E[\phi ,\rho ] &=&\displaystyle{\int d^DxH[\phi ]}  \nonumber \\
&=&\displaystyle{\sum_k\theta (-E_k[\phi ])E_k[\phi ]-\int d^Dx\phi (x)\rho
(x)+\frac 12\int d^Dxd^Dyv(x-y)\rho (x)\rho (y)}  \label{8}
\end{eqnarray}
where the boson field $\phi (x)$ is determined by the condition $\delta
E[\phi ,\rho ]/\delta \phi (x)=0$, which induces the equation, 
\begin{equation}
\sum_k\theta (-E_k[\phi ])\left( G_k(x)+\int d^Dy\phi (y)\frac{\delta G_k(y)%
}{\delta \phi (x)}\right) =\rho (x)  \label{9}
\end{equation}
where $G_k(x)=\int_0^1d\xi e^{2Q_k^R(x,\xi )}/\int d^Dxe^{2Q_k^R(x,\xi )}$,
and $Q_k(x,\xi )=Q_k^R(x,\xi )+iQ_k^I(x,\xi )$. This equation shows that the
boson field $\phi (x)$ is the functional of the electron density field $\rho
(x)$. The chemical potential is determined by the constraint equation, 
\begin{equation}
\sum_k\theta (-E_k[\phi ])=N  \label{10}
\end{equation}
where $N$ is the total electron number. With equation (\ref{9}), the ground
state energy can be represented as another form, which is only the
functional of the electron density field ($\phi (x)$ is the functional of $%
\rho (x)$ determined by (\ref{9})), 
\begin{equation}
E_g[\rho ]=E_0+\frac 12\int d^Dxd^Dy\left( v(x-y)\rho (x)\rho (y)-2\phi
(x)\phi (y)\sum_k\theta (-E_k[\phi ])\frac{\delta G_k(y)}{\delta \phi (x)}%
\right)  \label{11}
\end{equation}
where $E_0=\sum_k\theta (-E_k[\phi ])\epsilon _k$, and the self-energy can
be written as a simple form, $\Sigma _k[\phi ]=\int d^Dx\phi (x)G_k(x)$. The
equations (\ref{6}), (\ref{9}), (\ref{10}) and (\ref{11}) give the exact
expression of the ground state energy of the system as a functional of the
electron density, we can easily numerically calculate it and the ground
state electron density by taking $\delta E_g[\rho ]/\delta \rho (x)=0$.
These equations have more advantages than that in the Kohn-Sham scheme,
because that, 1). the expression of the ground state energy as the
functional of the electron density is exact, with it we can exactly
determine the ground state energy and the ground state electron density; 2).
with these equations, we can easily estimate the contributions from high
order terms, and obtain enough accurate results we hoped for some special
considerations; 3). Using these equations to calculate the ground state
energy and the ground state electron density, we may need much less computer
time than that in the Kohn-Sham scheme. The ground state wave-function of
the system can be obtained by the eigen-functionals, however, we cannot
write it as a simple form, because the boson field $Q_k(x,\xi =1)$ is the
function of the momentum $k$.

The expression of the ground state energy (\ref{11}) is valid not only for
Landau Fermi liquid (weak correlation systems), but also for non-Fermi
liquid (strongly correlated systems), because it is universal for weak and
strong electron interactions. Thus the Landau Fermi liquid and non-Fermi
liquid can be unifiably represented by our eigen-functional bosonization
theory, because the equation (\ref{11}) gives the physical properties of the
ground state, and in Ref.\cite{8} we show how to study the physical
properties of the excitation states, and how to calculate the correlation
functions of the systems. The phase field $Q_{k}(x,\xi)$ is a key parameter
for unifiably representing the Landau Fermi liquid and non-Fermi liquid. Its
imaginary part represents the electron correlation, and its real part ($%
D\geq 2$) only contributes to the ground state energy and the action of the
systems, which can be clearly seen in the expression of the ground state
energy (\ref{11}).

For a boson system, such as the liquid $^4He$, at zero temperature it has
Bose-Einstein condensation, and the bosons only occupy the state of the
momentum $k=0$, thus the equation (\ref{10}) is trivial, and the ground
state energy can be written as, 
\begin{equation}
E_g[\rho ]=\frac 12\int d^Dxd^Dy[v(x-y)\rho (x)\rho (y)-2\phi (x)\rho (x)]
\label{b1}
\end{equation}
where we have taken $\mu =\Sigma _0[\phi ]$. The equation (\ref{9}) reduces, 
\begin{equation}
G_0(x)+\int d^Dy\phi (y)\frac{\delta G_0(y)}{\delta \phi (x)}=\frac 1N\rho
(x)  \label{b2}
\end{equation}
It is noted that the last term in (\ref{b1}) is the contributions of kinetic
energy of the bosons, and in general it may be non-zero for interacting
boson systems. Due to the condensation of the bosons, we can easily obtain
the ground state wave-function of the boson systems, 
\begin{eqnarray}
\Psi (x_1,x_2,...,x_N) &=&\displaystyle{\ \left( \frac{A_0}{L^{D/2}}\right)
^N<\Psi (x_1,x_2,...,x_N,[\phi ])>_\phi }  \nonumber \\
\Psi (x_1,x_2,...,x_N,[\phi ]) &=&\displaystyle{\ e^{\sum_{i=1}^NQ_0(x_i,\xi
=1)}}  \label{b3}
\end{eqnarray}
where $<...>_\phi $ means the functional average over the boson field $\phi
(x)$. Using the method in Ref.\cite{8}, we can easily obtain the ground
state wave-function which is uniquely determined by single effective
potenial function, and has the expression very similar to usual correlated
basis functions\cite{16} that are the type of wave-function most often
employed in the study of the ground state properties of $^4He$. While the
functional $\Psi (x_1,x_2,...,x_N,[\phi ])$ is very similar to the
generalized London wave-function\cite{17}, where $f(x_i)=\exp
\{Q_0^R(x_i,\xi =1)\}$ and $S(x_i)=Q_0^I(x_i,\xi =1)$.

We now consider the electron motion under an external magnetic field ${\bf B}%
=(0,0,B)$, where the response of the system to the magnetic field cannot be
written as a simple density form. It is well-known that for enough strong
megnetic field, in low temperature limit a two-dimensional electron gas
shows the fractional quantum Hall effects due to the Coulomb interaction of
the electrons. Here we only give the exact expression of the ground state
energy of this system, and do not compare it with that obtained by
Laughlin's trial wave-functions\cite{14}, because that needs more detail
numerical calculations. Under the magnetic field, the Hamiltonian (\ref{2})
becomes, 
\begin{eqnarray}
H[\phi ,\rho ] &=&\displaystyle{\psi ^{\dagger }(x)\left( \frac 1{2m}(\hat{p}%
+\frac ec{\bf A})^2-\mu +\phi (x)\right) \psi (x)}  \nonumber \\
&-&\displaystyle{\phi (x)\rho (x)+\frac 12\int d^2yv(x-y)\rho (y)\rho (x)}
\label{12}
\end{eqnarray}
where the gauge field ${\bf A}=(-yB/2,xB/2,0)$, and for simplicity we take $%
D=2$. The eigen-functional equation corresponding to equation (\ref{3})
reads, 
\begin{equation}
\left( \frac 1{2m}(\hat{p}+\frac ec{\bf A})^2-\mu +\phi (x)\right) \Psi
_{nl}(x,[\phi ])=E_{nl}[\phi ]\Psi _{nl}(x,[\phi ])  \label{13}
\end{equation}
and the eigen-values are 
\begin{eqnarray}
E_{nl}[\phi ] &=&\epsilon _n+\Sigma _{nl}[\phi ]  \nonumber \\
\Sigma _{nl}[\phi ] &=&\displaystyle\int_0^1{d\xi }\int {d^2x\phi (x)|\Psi
_{nl}(x,[\xi \phi ])|^2}  \label{14}
\end{eqnarray}
where $\epsilon _n=\hbar \omega _0(n+1/2)-\mu $, $n=0,1,2,...$, and $\omega
_0=eB/(mc)$ is the cyclotron frequency. It is well-known that at $\phi (x)=0$
the eigen-equation (\ref{13}) has exact solutions, $\Psi _{nl}(x,[0])=\psi
_{nl}(x)=a_{nl}(z-\partial _{z^{*}})^l(z^{*}-\partial _z)^n\exp \{-zz^{*}\}$%
, $l=0,1,2,..,L_{max}=\Phi /\Phi _0$, where $\Phi =BS$ is the total flux, $%
\Phi _0=2\pi \hbar c/e$ the flux quantum, $z=(x-iy)/(2l_B)$, $%
z^{*}=(x+iy)/(2l_B)$, $l_B=(\hbar c/(eB))^{1/2}$ is the magnetic length, and 
$a_{nl}$ a normalization constant. Therefore, the eigen-functionals have the
following form, 
\begin{equation}
\Psi _{nl}(x,[\xi \phi ])=A_{nl}\psi _{nl}(x)e^{Q_{nl}(x,\xi )}  \label{15}
\end{equation}
where $A_{nl}$ is a normalization constant, and the phase field $%
Q_{nl}(x,\xi )$ satisfies Eikonal-type equation, 
\begin{equation}
\left( \frac{\hat{p}^2}{2m}+(\frac e{mc}{\bf A}+{\bf a}_{nl})\cdot \hat{p}%
\right) Q_{nl}(x,\xi )+\frac{[\hat{p}Q_{nl}(x,\xi )]^2}{2m}+\xi \phi (x)=0
\label{16}
\end{equation}
where ${\bf a}_{nl}=(1/m)\hat{p}\ln \psi _{nl}(x)$.

Following the above same procedures, we have the exact expression of the
ground state energy as the electron density field $\rho(x)$, 
\begin{equation}
E_{g}[\rho]=E_{0}+\frac{1}{2}\int d^{2}x d^{2}y \left( v(x-y)\rho(x)\rho(y)-
2\phi(x)\phi(y)\sum_{nl}\theta(-E_{nl}[\phi])\frac{\delta G_{nl}(y)}{\delta
\phi(x)}\right)  \label{17}
\end{equation}
where $E_{0}=\sum_{nl}\theta(-E_{nl}[\phi])\epsilon_{n}$, and the boson
field $\phi(x)$ is the function of the electron density $\rho(x)$, and is
determined by the equation, 
\begin{equation}
\sum_{nl}\theta(-E_{nl}[\phi])\left( G_{nl}(x)+\int d^{2}y\phi(y) \frac{%
\delta G_{nl}(y)}{\delta\phi(x)}\right)=\rho(x)  \label{18}
\end{equation}
where $G_{nl}(x)=\int^{1}_{0} d\xi |\psi_{nl}(x)|^{2}e^{2Q^{R}_{nl}(x,\xi)}
/\int d^{2}x |\psi_{nl}(x)|^{2}e^{2Q^{R}_{nl}(x,\xi)}$, and the self-energy
can written as a simple form, $\Sigma_{nl}[\phi]=\int d^{2}x
\phi(x)G_{nl}(x) $. The chemical potential is determined by the constraint
condition, 
\begin{equation}
\sum_{nl}\theta(-E_{nl}[\phi])=N  \label{19}
\end{equation}
where $N$ is the total electron number. The equations (\ref{16}), (\ref{17}%
), (\ref{18}) and (\ref{19}) can be used to exactly determine the ground
state energy and the ground state electron density of the system by taking $%
\delta E_{g}[\rho]/\delta\rho(x)=0$. In the low temperature limit, for
enough strong magnetic field, the system shows the fractional quantum Hall
effects for some special filling factors $\nu=N\Phi_{0}/\Phi$. The
expression of the ground state energy (\ref{17}) has a great advantages
comparing with the Laughlin's trial wave-functions: 1). it is a microscopic
theory expression. 2). it shows that the odd- and even-denominator's
fractional quantum Hall states have the same expression of the ground state
energy. 3). it can exactly determine the ground state electron density of
the system. 4). it is very simple to study the fractional quantum Hall
effects in higher Landau levels ($n\geq 1$).

However, the ground state wave-function of the system cannot be written as a
simple form even for the lowest Landau level ($n=0$), because the phase
field $Q_{nl}(x,\xi=1)$ depends on the quantum numbers $n$ and $l$. In the
lowest Landau level, if we approximately take $Q_{0l}(x,\xi=1)\sim Q_{00}(x)$%
, we can obtain the following expression of the ground state wave-function, 
\begin{equation}
\Psi(x_{1},x_{2},...,x_{N})=A\prod^{N}_{i=1}z_{i}\prod_{i>j}(z_{i}-z_{j})
e^{\sum^{N}_{i=1}|z_{i}|^{2}}<e^{ \sum^{N}_{i=1}Q_{00}(x_{i})}>_{\phi}
\label{20}
\end{equation}
where $A$ is a normalization constant, and the factor $\prod_{i>j}(z_{i}-
z_{j})$ guarantees the anti-commutation of the electrons. The last factor $%
<\exp\{\sum^{N}_{i=1}Q_{00}(x_{i})\}>_{\phi}$ is very similar to the ground
state wave-function of the boson systems (\ref{b3}), and it is very clear
that this factor is the contribution of the electron interactions. This
ground state wave-function is different from the Laughlin's trial
wave-functions of the odd-dinominator fractional quantum Hall states. We
believe that with the equations of the ground state energy and the ground
state wave-function (\ref{17}) and (\ref{20}), respectively, we can obtain
more important informations of the fractional quantum Hall effects than that
by the Laughlin's trial wave-functions, because they are directly derived
from the microscopic theory.

In summary, by introducing the Lagrangian multiplier (boson field) which
makes the Hamiltonian only have the quadratic form of the electron/boson
operators, and solving the eigen-functional equation of the
electrons/bosons, we have obtained the exact expressions of the ground state
energy as the functional of the electron/boson density for interacting
electron systems, interacting boson systems, and the two-dimensional
electron gas under an external magnetic field, respectively. By taking $%
\delta E_g[\rho ]/\delta \rho (x)=0$, we can exactly determine the ground
state energy and the ground state electron/boson density of the quantum
many-particle systems. With the present method and the eigen-functional
bosonization theory\cite{8} which can exactly calculate the action and the
correlation functions of the quantum many-particle systems, we can establish
an unified theory of the quantum many-particle systems, which is valid not
only for weakly correlated fermion/boson systems, but also for strongly
correlated fermion/boson systems. The key points of this unified theory are
that: 1). we introduce the Lagrangian multiplier (boson field) $\phi (x)$
which takes the particle density $\rho (x)=\psi ^{\dagger }(x)\psi (x)$ as a
constraint condition, so that the Hamiltonian can only have the quadratic
form of the particle (fermion/boson) operators. 2). by introducing phase
field $Q_k(x,\xi )$ which is a functional of the boson field $\phi (x)$, we
solve the eigen-functional equation of the particles, so that we can use the
phase field to completely represent the kinetic energy of the systems. 3).
the phase field $Q_k(x,\xi )$ is a key parameter hidden in the quantum
many-particle systems, its imaginary part represents the particle
correlation strength, and its real part only contributes to the ground state
energy and action of the systems. 4). we are able to use the particle
density field $\rho (x)$ to exactly represent the ground state energy and
action of the quantum many-particle systems.

\end{document}